\documentclass[pra,aps,twocolumn,superscriptaddress,showpacs]{revtex4}

\usepackage{amsfonts,amsmath,bm}
\usepackage[OT4]{fontenc}
\usepackage{color}
\usepackage{graphicx}
\usepackage{epstopdf}
\usepackage{color}

\usepackage{hyperref}

\hypersetup{
   colorlinks=true,         
   linkcolor=blue,          
   citecolor=blue,           
   urlcolor=Violet          
}

\newcommand{\ket}[1]{\ensuremath{|#1\rangle}}
\newcommand{\braket}[2]{\ensuremath{\left\langle#1 \vphantom{#2}\middle|  #2 \vphantom{#1}\right\rangle}}
\newcommand{\ketbra}[2]{\ensuremath{|#1\rangle\!\langle#2|}}

\newcommand{\tr}[1]{\mathrm{Tr}\left( #1 \right)}

\begin{document}

\title{Quantum-state transfer in spin chains via isolated resonance of terminal spins} 

\author{Kamil Korzekwa}
\affiliation{Department of Physics, Imperial College London, London SW7 2AZ, United Kingdom}
\affiliation{Institute of Physics, Wroc{\l}aw University of Technology, 50-370 Wroc{\l}aw, Poland}
\author{Pawe{\l} Machnikowski} 
\affiliation{Institute of Physics, Wroc{\l}aw University of Technology, 50-370 Wroc{\l}aw, Poland}
\author{Pawe{\l} Horodecki}
\affiliation{Faculty of Applied Physics and Mathematics, Technical University of Gda\'{n}sk, PL-80-952 Gda\'{n}sk, Poland}
\affiliation{National Quantum Information Centre of Gda\'{n}sk, PL-81-824 Sopot, Poland}

\begin{abstract}

We propose a quantum-state transfer protocol in a spin chain that requires only the control of the spins at the ends of the quantum wire. The protocol is to a large extent insensitive to inhomogeneity caused by local magnetic fields and perturbation of exchange couplings. Moreover, apart from the free evolution regime, it allows one to induce an adiabatic spin transfer, which provides the possibility of performing the transfer on demand. We also show that the amount of information leaking into the central part of the chain is small throughout the whole transfer process (which protects the information sent from being eavesdropped) and can be controlled by the magnitude of the external magnetic field.

\end{abstract}


\maketitle

\section{Introduction} 
\label{sec:intro}

For practical quantum information processing, one needs not only controllable and decoherence-resistant qubits but also short-range communication channels providing inter-qubit communication during more complex computation processes \cite{kielpinski02}. Chains of coupled qubits, referred to as \textit{spin chains}, have been proposed for such channels \cite{bose03,bose07a}, since they do not require conversion between different physical encodings and therefore present an interesting alternative to photonic flying qubits \cite{gheri98,blinov04}. This approach has recently been extended to include the transfer of higher-dimensional states \cite{qin13} and the use of qutrit (spin-1) chains \cite{wiesniak13}.

A useful information channel must work with high fidelity. Strictly perfect transfer is possible in chains with individually tuned $XY$ couplings \cite{christandl04,nikolopoulos04a,nikolopoulos04b,camposvenuti07,difranco08,wang11a}, which may assure commensurability of the spectrum \cite{albanese04}. However, this is not admitted by chains coupled with isotropic Heisenberg interactions \cite{wiesniak08} (which are typical for spins confined in quantum dots \cite{loss98}) and an arbitrarily perfect state transfer can only be achieved for particular lengths of the spin chain \cite{godsil12}. As an alternative, applying a magnetic field with a simple spatial variation along the chain has been proposed \cite{shi05,gianluca13}. It is also possible to assure high-fidelity transfer by appropriately switching the couplings  \cite{eckert07} in order to induce adiabatic transfer by applying global \cite{fitzsimons06} or local \cite{balachandran08,murphy10} external fields. The transfer fidelity can also be improved by using a receiver with memory \cite{giovannetti06} or by multiqubit encoding \cite{osborne04,wang09,markiewicz09,bishop10}.

While achieving specific control over the whole spin chain may be quite demanding, protocols involving only local manipulation seem to be feasible. Such proposals based on time-dependent control methods include transfer procedures involving logical gates on the terminal bits \cite{burgarth07,difranco08,burgarth10}, local switching of a perturbation \cite{schirmer09}, and controlling the interactions between pairs of terminal spins \cite{haselgrove05,zwick13}. Among time-independent control methods the most notable are based on weak coupling of the terminal (sender and receiver) qubits to a spin chain \cite{wojcik05,plenio05,camposvenuti07,wojcik07,oh11,ajoy13} and on the use of strong local magnetic field on two nodes of the chain \cite{plastina07,casaccino09,linneweber12,lorenzo13,paganelli13}.

Clearly, the feasibility of a transfer protocol depends on the required degree of control over the spin chain. From this point of view, the solutions involving precisely engineered couplings along the chain or local time-dependent control are less favorable than those based on controlling only the terminal qubits. Another desirable possibility is a conclusive transfer, such that the receiving party can be sure that the transfer has been completed. In order to achieve this, dual rail protocols have been proposed \cite{burgarth05a}. An information transfer protocol should also be resilient to environmental noise \cite{burgarth06a,hu10,jeske13} and to imperfections and fluctuations of the chain parameters \cite{yang10,burrell09,ronke11,zwick11,bruderer12}.

In our previous work \cite{korzekwa11} we proposed a quantum dot implementation of a quantum state transfer channel that supported high-fidelity transfer of the state of a terminal dot. In this paper we generalize and develop our idea to propose a spin-chain-based protocol that requires only the control of the local magnetic field on terminal qubits. Compared with previously proposed protocols based on similar controlling scheme, our protocol allows for almost perfect transfer not only along uniformly coupled chains but also along randomly coupled ones. This can be achieved by ``compensating'' the effect of fluctuations in the couplings through the control of the relative strength of magnetic field on two terminal nodes. Moreover, we show that our protocol provides protection against eavesdropping through low leakage of quantum information to the central part of the chain available to the third party. Finally, apart from the free dynamics, it admits an adiabatic evolution in which the qubit state is transferred from one terminal to the other by sweeping the energy of one of the terminal states through resonance by a local field applied on either end of the chain. The main advantage of this proposal over other time-dependent transfer protocols lies in its simplicity: It only requires a magnetic field sweep, which is independent of the parameters of the spin chain used. It is also worth noting that transfer on demand can be obtained either by the sender or by the receiver, which compares favorably with some of the proposed schemes, where both parties must be involved in the transfer process.

The paper is organized as follows. In Sec. \ref{sec:model}, we describe the system and the methods of simulation used. Next, in Sec. \ref{sec:transfer}, we discuss two versions of our protocol: free evolution and adiabatic transition driven by magnetic field sweep at the terminal sites of the chain. In Sec. \ref{sec:inf} we study the leakage of quantum information from the terminal nodes of the chain (controlled by communicating parties) to the central ones (which may be available to eavesdropper). Finally, Sec. \ref{sec:concl} concludes the paper.

\section{Model and method} 
\label{sec:model}

We consider a model consisting of $N$ linearly arranged spins with the nearest neighbor XY couplings $J_k$ and a local magnetic field at each spin site $B_k$. The model is restricted to at most one spin-flip excitation in the chain. Then the Hamiltonian of the system is given by
\begin{equation}
\label{eq:hamiltonian}
    H=\sum_{k=1}^{N-1} J_k\left(|k\rangle\!\langle k+1| +\mbox{H.c.}\right)+\sum_{k=1}^N B_k|k\rangle\!\langle k|,
\end{equation} 
where $|k\rangle$ denotes the state with an inverted (excited) spin at the $k$th site of the chain. We assume that only the terminal spins ($k=1, N$) are controlled by external fields $B_k^{\mbox{\scriptsize(ext)}}$, while the fields on all the nodes (including the terminal ones) are random (but time independent), reflecting the effects of an inhomogeneous environment. We include the inhomogeneity of the spin chain by using the normal distribution for both the local magnetic fields, $B_k$, and the couplings, $J_k$, with the standard deviations $\sigma_B$ and $\sigma_J$, respectively. The average value of $J_{k}$ is fixed and equal to $J$ for all $k$ and the the average value of the local field, $\bar{B}_k$, is nonzero (and equal to the applied control field) only at the terminal ends. 

In this paper, we present two possible kinds of the state transfer protocol. In the first case the free transfer with time-independent external magnetic fields is assumed, modeled by numerical diagonalization of the Hamiltonian and finding the evolution in the basis of eigenstates. In the second case the transfer takes place adiabatically with time-dependent external field and is simulated numerically to find the numerical solution of the Schr\"odinger equation.

\section{Quantum state transfer} 
\label{sec:transfer}

We consider a generic initial spin state localized at the first (sender's) node of the chain,
\begin{equation}
\label{eq:state_initial}
\ket{\Psi(0)}=\cos\frac{\theta}{2}\ket{0}+e^{i\phi}\sin\frac{\theta}{2}\ket{1},
\end{equation}
where $\ket{0}$ denotes state of the chain with no spins inverted. After time $t$ the system state evolves into 
\begin{equation}
\label{eq:state_evolved}
\ket{\Psi(t)}=\cos\frac{\theta}{2}\ket{0}+e^{i\phi}\sin\frac{\theta}{2}\sum_{k=1}^Nc_k\ket{k},
\end{equation}
and we are interested in the fidelity between the initial sender's spin and the final receiver's $N$th spin. Due to the trivial evolution of the state $\ket{0}$ [it is an eigenstate of the Hamiltonian, Eq. \eqref{eq:hamiltonian}, with zero energy] one may consider only the transfer fidelity of the initial state \mbox{$|\Psi(0)\rangle=|1\rangle$}, 
\begin{equation*}
F=|\braket{N}{\Psi(t)}|^2,
\end{equation*} 
as the transfer fidelity of a generic state is a direct function of $F$. In fact, it is easy to show that the fidelity averaged over all possible initial states of the first spin (over the whole Bloch sphere) is given by \mbox{$F_{\mathrm{avg}}=\frac{1}{3}+\frac{1}{6}(1+F)^2$}.

Let us also note that, for the spin chain under consideration, the transfer of entanglement between the sender's spin and a spin isolated from the rest of the chain to the receiver's spin depends directly on the state transfer fidelity. Using the Wootters formula \cite{wootters98}, one finds that for an initial singlet state (maximally entangled state of the sender's and isolated spins) the entanglement of formation between the isolated spin and the receiver's one is given by
\begin{equation*}
E[\rho_{0N}]=-x_+\log_2x_+-x_-\log_2x_-,
\end{equation*}
with $x_{\pm}=(1\pm\sqrt{1-F})/2$.

\subsection{Exploting isolated resonance}

Due to the external magnetic field applied at the terminal sites, the energies of the spin-up states localized on those sites are higher than for those localized inside the chain. If this external magnetic field is strong enough ($\gtrsim 2J$), it results in the formation of two eigenstates $|\Psi_N\rangle$ and $|\Psi_{N-1}\rangle$ that are energetically separated from all the others and mainly composed of the states $|1\rangle$ and $|N\rangle$. Due to indirect coupling via the spin chain, these two eigenstates form an anticrossing when the on-site energies of the states $|1\rangle$ and $|N\rangle$ are brought to resonance. While the position of this resonance as a function of $\delta B=B^{\mbox{\scriptsize(ext)}}_1-B^{\mbox{\scriptsize(ext)}}_N$ is randomly shifted in the presence of parameter fluctuations in the central part of the chain (due to coupling-induced energy shifts), its width remains almost constant \cite{korzekwa11}.
The two states can be brought to resonance by applying an asymmetric external magnetic field $B^{\mbox{\scriptsize(ext)}}_1=B^{\mbox{\scriptsize(ext)}}_N+\delta B$ which compensates the shift of the resonance and leads to the resonantly enhanced tunneling \cite{livescu89,vandervaart95} (we will refer to this situation as ``compensated inhomogeneous chain''). Close to resonance we can restrict, to a good approximation, the free evolution of the system to the subspace $\{|\Psi_N\rangle,|\Psi_{N-1}\rangle\}$. Then the evolution corresponds to the oscillation of the spin-inverted state between the terminal sites of the chain. In the simulations described in the following sections we set $\bar{B}_N/J=5$.

\begin{figure}[tb]
\begin{center}
\includegraphics[width=\columnwidth]{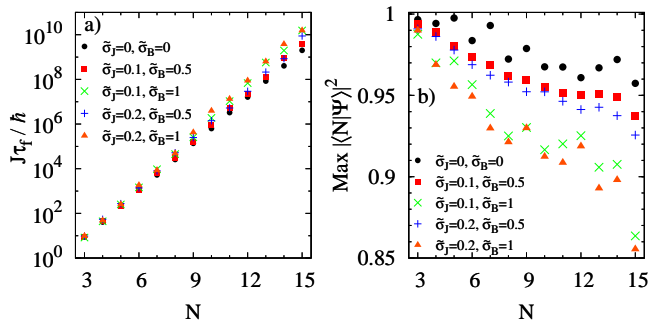}
\end{center}
\caption{\label{fig:transfer_time} Free evolution in homogeneous (circles) and compensated inhomogeneous chains for the values of $\tilde{\sigma}_J$ and $\tilde{\sigma}_B$ as shown. The data for inhomogeneous chains are averaged over 100 random realizations. (a) Transfer time as a function of the chain length; (b) maximum achieved fidelity as a function of the chain length.}
\end{figure} 

\subsection{Free transfer}
\label{sec:free}

The oscillation period is determined by the width of the resonance $2V$, which critically depends on $N$. This is reflected in the free transfer time dependence \mbox{$\tau_f=\pi\hbar/(2V)$}. In Fig.~\ref{fig:transfer_time} we show dependence of $\tau_f$ and maximum obtained fidelity on $N$ for a few values of the normalized standard deviations $\tilde{\sigma}_J=\sigma_J/J$ and $\tilde{\sigma}_B=\sigma_B/J$. As long as the inhomogeneity is not too strong, it weakly affects the width of the resonance and the free transfer time for a compensated inhomogenous chain is of the order of the free transfer time in the homogenous chain [Fig.~\ref{fig:transfer_time}(a)]. Since the fidelity depends mainly on the degree of localization of $|\Psi_N\rangle$ and $|\Psi_{N-1}\rangle$ on the terminal sites ($|1\rangle$ and $|N\rangle$), it is well characterized by the parameter $\Delta=1-|\langle\ 1|\Psi_N\rangle|^2-|\langle 1|\Psi_{N-1}\rangle|^2$. In our model, $\Delta$ does not depend on the chain length $N$, so the transfer fidelity should also weakly depend on $N$. This is in fact confirmed by our simulations [Fig.~\ref{fig:transfer_time}(b)]. We have observed also that the inhomogeneity of the magnetic field affects both transfer time and achieved fidelity stronger than the inhomogeneity of the couplings.

\begin{figure}[tb]
\begin{center}
\includegraphics[width=\columnwidth]{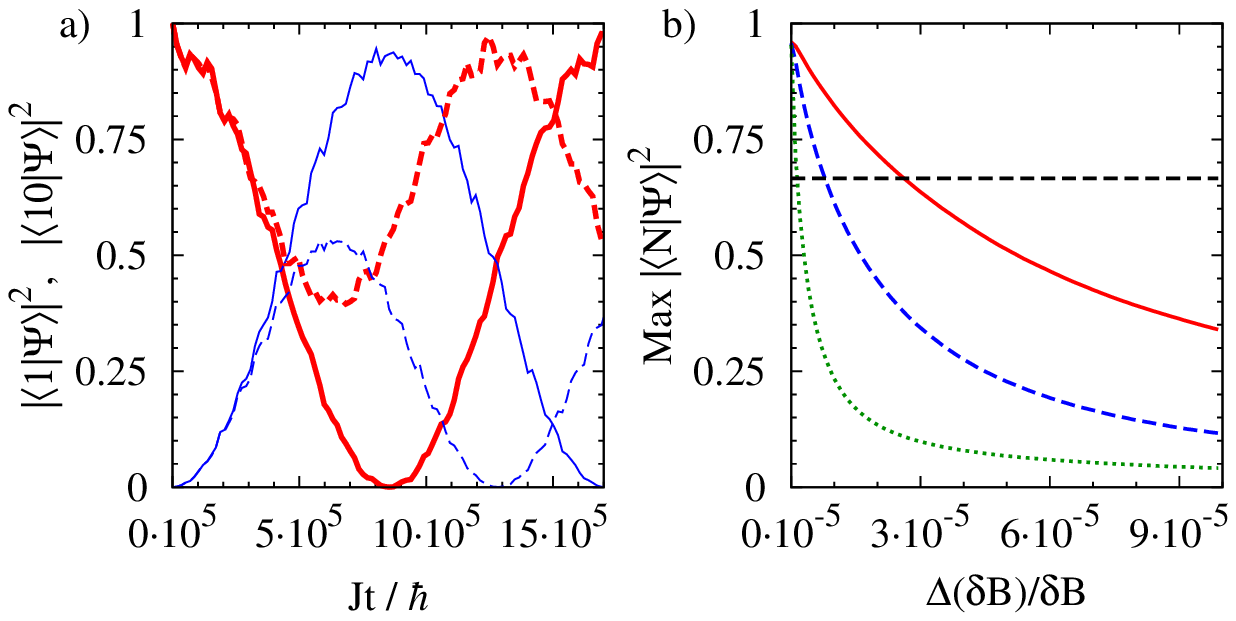}
\end{center}
\caption{\label{fig:free_transfer} (a) Occupation probabilities for the initial (thick red lines) and final (thin blue lines) nodes for a compensated inhomogenous spin chain with \mbox{$N=10$}, \mbox{$\tilde{\sigma}_J^2=0.1$}, \mbox{$\tilde{\sigma}_B^2=0.5$}. Solid lines: Perfect compensation; dashed lines: $0.002\%$ deviation from perfect compensation. (b) Maximum obtained fidelity (averaged over 100 random realizations for \mbox{$\tilde{\sigma}_J^2=0.1$}, \mbox{$\tilde{\sigma}_B^2=0.5$}) as a function of the deviation from the perfectly compensating magnetic field for an inhomogenous chain of length \mbox{$N=9$} (red solid line), \mbox{$N=10$} (blue dashed line) and \mbox{$N=11$} (green dotted line). Horizontal black dashed line shows the classical limit of efficiency of quantum transfer, \mbox{$F=2/3$} \cite{bose07a}.}
\end{figure}

In Fig. \ref{fig:free_transfer}(a) we present an exemplary evolution of the spin-up state in a compensated inhomogenous chains and show that the transfer fidelity $F$ achieved for the perfect compensation is very high (over $0.95$). However, since the weak indirect coupling between the two terminal states leads to a very narow resonance, $F$ is very sensitive to small variations of the compensating field, i.e., a small deviation of magnetic field results in a big decrease of the maximum fidelity obtained [Fig.~\ref{fig:free_transfer}(b)].

\begin{figure}[tb]
\begin{center}
\includegraphics[width=\columnwidth]{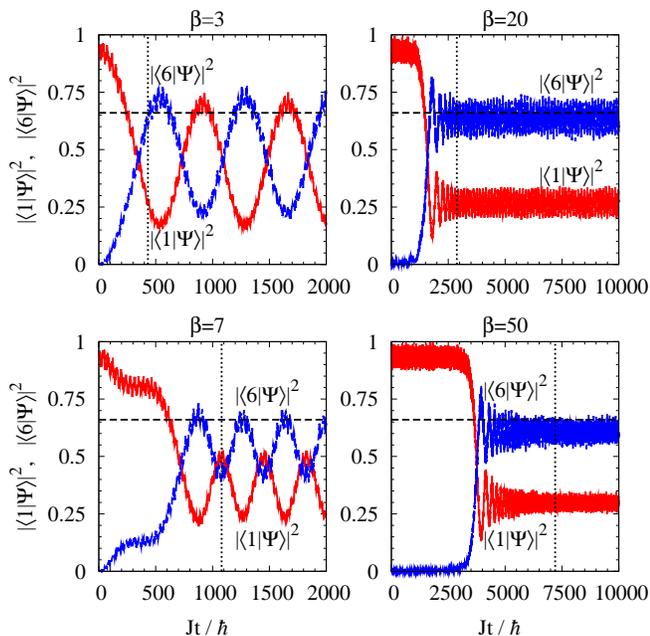}
\end{center}
\caption{\label{fig:adiabatic_transfer_beta} Occupation probabilities for the initial (red lines) and final (blue lines) nodes for an inhomogenous spin chain with \mbox{$N=6$}, \mbox{$\tilde{\sigma}_J^2=0.1$}, \mbox{$\tilde{\sigma}_B^2=0.5$} during adiabatic evolution. Horizontal dashed lines show the desired fidelity \mbox{$F=0.66$} obtained from the Landau Zener formula, and vertical dotted lines indicate the end of the magnetic field sweep.} 
\end{figure}

\subsection{Adiabatic transfer}
\label{sec:adiabatic}

The requirement of a very precise control of the magnetic field makes it a rather demanding task to obtain a free transfer. Moreover, in the presence of randomness, the optimal transfer time is unpredictable. To overcome this problem, we propose an adiabatic variation of our protocol. By slowly changing the asymmetry of the external magnetic field $\delta B$ one can sweep the energy levels of $|\Psi_N\rangle$ and $|\Psi_{N-1}\rangle$ through the resonance. It is worth noting that this procedure can be done either by the sender or by the receiver. The system evolution in this case can be described by an effective two-level model including the states $|\Psi_N\rangle$ and $|\Psi_{N-1}\rangle$, with the coupling $V$ taken as half of the energy splitting at the resonance. Then, using the Landau-Zener formula \cite{wittig05} for nonadiabatic transition probabilities,
\begin{equation}
    P_{\mathrm{na}}=\exp\left(-\frac{2\pi}{\hbar}\frac{|V|^2}{\alpha}\right),
\end{equation}
we found the dependence of the speed of magnetic field sweep $\alpha=dB/dt$ as a function of the desired fidelity $F=1-P_{\mathrm{na}}$. 

Formally, the field is swept over ($-\infty,\infty$), which is unrealistic. To achieve a finite transfer time we narrow the limits of the magnetic field sweep to the area where the energy separation of the states is smaller than $\beta V$ for a certain parameter $\beta$ (we assume that interaction is negligible for $\Delta E>\beta V$). In this way we obtain the adiabatic transfer time $\tau_{\mathrm{a}}=-[\hbar\beta/(\pi V)]\ln\left(1-F\right)$, whose ratio to the free transfer time $\tau_f$ (for a given $\beta$) depends only on the desired fidelity, $\tau_{\mathrm{a}}/\tau_{{\mathrm{f}}}=-2\beta/\pi^2\ln\left(1-F\right)$.

\begin{figure}[tb]
\begin{center}
\includegraphics[width=\columnwidth]{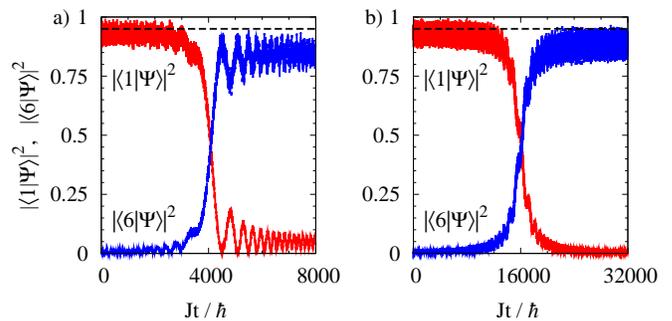}
\end{center}
\caption{\label{fig:adiabatic_transfer_alpha} Occupation probabilities for the initial (red lines) and final (blue lines) nodes for an inhomogenous spin chain with \mbox{$N=6$}, \mbox{$\tilde{\sigma}_J^2=0.1$}, \mbox{$\tilde{\sigma}_B^2=0.5$} during an adiabatic evolution for $\beta=20$. Horizontal dashed lines show the desired fidelity \mbox{$F=0.95$}. (a) Speed of the magnetic field sweep $\alpha$ calculated from the Landau-Zener formula; (b) 4-times-slower magnetic field sweep $0.25\alpha$.} 
\end{figure}

To confirm the agreement of the effective two-level model with the Landau-Zener result we simulated the evolution of the full system, where we swept \mbox{$-\beta V<\delta B<\beta V$} with a constant speed $\alpha$. For large enough values of $\beta$, the adiabatic evolution stabilizes (further increasing of $\beta$ does not change the evolution) and we get a good agreement between the Landau-Zener and simulated results (Fig.~\ref{fig:adiabatic_transfer_beta}). However, for large $F$, the actual fidelity is lower than expected from the Landau-Zener formula [Fig.~\ref{fig:adiabatic_transfer_alpha}(a)], even for a relatively large value of $\beta$ (further increase of this parameter does not bring a considerable improvement). It remains true also for a decreased speed of the magnetic field sweep [Fig.~\ref{fig:adiabatic_transfer_alpha}(b)]. This can be explained by noting that the total occupation probability of the terminal nodes at the end of the evolution is lower than $1$, which in turn indicates that the origin of the problem with achieving very high fidelity lies in the leakage to the nodes inside the chain. 

\section{Leakage of quantum information}
\label{sec:inf}

\begin{figure}[t] \begin{minipage}[t]{\columnwidth}
\begin{center}
\includegraphics[width=\columnwidth]{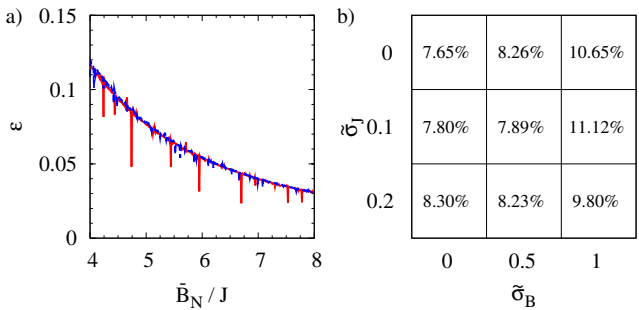}
 \end{center} \end{minipage}
\caption{\label{fig:Central_Inf} Total occupation probability of the nodes in the central part of the chain, $\epsilon$, averaged over the free transfer time. (a) Dependance of $\epsilon$ on the normalized external magnetic field \mbox{$\bar{B}_N/J$} for a homogeneous chain of length \mbox{$N=5$} (red solid line) and \mbox{$N=15$} (blue dashed line); (b) $\epsilon$ for different inhomogeneity parameters $\sigma_J^2$, $\sigma_B^2$ averaged over 100 random realizations of a chain of length \mbox{$N=10$}.} 
\end{figure}

To estimate the amount of leaking information let us define 
\begin{equation}
\epsilon=1-|c_1|^2-|c_N|^2,
\end{equation}
i.e., the total occupation probability for all the states $|k\rangle,2\leq k\leq N-1$ [coefficients $c_k$ as defined in Eq. \eqref{eq:state_evolved}]. In both cases, adiabatic and free transfer, simulations show that $\epsilon$ remains very small throughout the whole evolution (around $10\%$) and is mainly determined by the magnitude of the external magnetic field responsible for isolating terminal states [Fig.~\ref{fig:Central_Inf}(a)]. Moreover it is observed that the value of $\epsilon$ does not strongly depend on the chain length or the inhomogeneity parameters $\sigma_J^2$, $\sigma_B^2$ [Fig.~\ref{fig:Central_Inf}(b)].

The low value of $\epsilon$ not only allows for obtaining high fidelity transfer but also is important from the point of view of protection against eavesdropping. In order to show this, let us consider the following scenario. The sender prepares one of the two pure states from an arbitrarily chosen basis, i.e., a generic qubit state defined in Eq. \eqref{eq:state_initial} or a state orthogonal to it. Now the essential question is as follows: How well can the eavesdropper distinguish which of the two states have been sent? In order to answer this question one may define a measure of the information available to the eavesdropper as 
\begin{equation}
D:=\frac{1}{2}||\rho-\rho^{\perp}||_1=\frac{1}{2}\tr{\left|\rho-\rho^{\perp}\right|},
\end{equation}
where $\rho$ and $\rho^{\perp}$ are the reduced density matrices of the part of the chain that eavesdropper has access to and that correspond to two orthogonal initial sender's states. Such a choice of measure is justified by its operational meaning, as $(1+D)/2$ is the average success probability for distinguishing between the states $\rho$ and $\rho^{\perp}$ when they are equally probable to occur \cite{helstrom76} (see also Ref. \cite{fuchs99}). 

We will focus on two limiting types of eavesdroppers: when the third party has access to all but terminal spins (which we will refer to as the powerful eavesdropper) and when this access is limited only to a single $n$th node inside the chain (weak eavesdropper). In the first case the eavesdropper's reduced states are given by
\begin{subequations}
\begin{eqnarray}
\rho&=&\left(1-\epsilon\sin^2\frac{\theta}{2}\right)\ketbra{\tilde{0}}{\tilde{0}}+\epsilon\sin^2\frac{\theta}{2}\ketbra{\tilde{\Psi}}{\tilde{\Psi}}\nonumber\\
&&+\cos\frac{\theta}{2}\sin\frac{\theta}{2}\left(\sqrt{\epsilon}e^{i\phi}\ketbra{\tilde{\Psi}}{\tilde{0}}+\mathrm{h.c.}\right),\label{eq:reduced1}\\
\rho^{\perp}&=&\left(1-\epsilon\cos^2\frac{\theta}{2}\right)\ketbra{\tilde{0}}{\tilde{0}}+\epsilon\cos^2\frac{\theta}{2}\ketbra{\tilde{\Psi}}{\tilde{\Psi}}\nonumber\\
&&-\cos\frac{\theta}{2}\sin\frac{\theta}{2}\left(\sqrt{\epsilon}e^{i\phi}\ketbra{\tilde{\Psi}}{\tilde{0}}+\mathrm{h.c.}\right)\label{eq:reduced2},
\end{eqnarray}
\end{subequations}
where tildes indicate states of the system without terminal spins, i.e., states of the central part of the chain of length $N-2$, and
\begin{equation*}
\ket{\tilde{\Psi}}=\left(\sum_{k=2}^{N-1}c_k\ket{\tilde{k}}\right)\Big/\left(\sum_{k=2}^{N-1}|c_k|^2\right)=\frac{1}{\epsilon}\sum_{k=2}^{N-1}c_k\ket{\tilde{k}}.
\end{equation*}
For the weak eavesdropper expressions for reduced states are very similar to the ones given by Eqs. \eqref{eq:reduced1} and \eqref{eq:reduced2}: one simply needs to replace $\epsilon$ with $|c_n|^2$, $\ket{\tilde{0}}$ with $\ket{0}_n$, and $\ket{\tilde{\Psi}}$ with $\ket{1}_n$ (where $\ket{0}_n$ and $\ket{1}_n$ denote the ground and excited state of the $n$th spin, respectively). In both cases the information available to the third party is given by
\begin{equation}
D=\sqrt{a^2\cos^2\theta+a\sin^2\theta},
\end{equation}
where $a=\epsilon$ for the powerful eavesddropper and $a=|c_n|^2$ for the weak one. As $|c_n|^2\leq\epsilon$ the low value of $\epsilon$ throughout the whole evolution provides protection against both considered kinds of eavesdroppers. Note that $D$ is minimized for states sent from the computational basis, as a consequence of the form of the considered Hamiltonian (for which $\ket{0}$ is an eigenstate).

Let us also mention one more issue connected with the low and controllable value of $\epsilon$ in our protocol. Due to the nearest-neighbor couplings and the continuity equation for probability, the rate at which the occupation probability of the receiver's spin changes (so also the rate of the state transfer) depends on the occupation probability of the spins in the central part of the chain, i.e., on $\epsilon$. Therefore, increasing the transfer fidelity and the protection against eavesdropping by decreasing $\epsilon$ slows down the rate at which the state is transferred. This not only means that the receiver has to wait longer to get the state, but also increases the time needed for the eavesdropper to intercept the quantum state transmitted through the spin chain. 

Having no time limitation, the information can be completely intercepted with the use of the free transfer protocol. The eavesdropper can simply couple his external spin, $S_{\mathrm{E}}$, to the one inside the chain and, using the magnetic field, tune its energy to the resonance with the sender's spin. As the coupled external spin affects the energy spectrum of the whole chain, it decouples the original sender's and receiver's spins by shifting their energy from the resonance. This results in the formation of a new, effectively two-level, subsystem, where $S_{\mathrm{E}}$ plays the role of the receiver's spin. Generally, the time needed to intercept information can be decreased by increasing the occupation probability of the spins in the central part of the chain during the evolution. The powerful eavesdropper can do this by applying constant magnetic field on all spin sites inside the chain, thus effectively reducing the energetic separation of terminal spins, which results in an increased localization of the excitation inside the chain [see Fig.~\ref{fig:Central_Inf}(a)]. This clearly suggests that there must be a trade-off involving the duration of the eavesdropping, the energy available to the eavesdropper and the amount of information intercepted. This question is, however, beyond the scope of the present paper and remains an interesting problem for further study.

As a final remark let us note that, although spin chains are only suggested as short-distance communication devices, so eavesdropping is usually not considered a concern, the discussion presented in this section gives a description of the general behavior of quantum information in the investigated system. Using it one can, for example, get some insight into the destructive effect of the environment on the state transfer. To see this, let us consider the situation when the control of the sender and receiver over terminal spins allows them to protect these nodes against environmental effects, so only the central part of the chain couples to external degrees of freedom. Now, as the personalized eavesdropper cannot distinguish between two orthogonal states being sent, neither can the environment. Therefore, one can expect that the correlations between the spin chain and its environment will develop more slowly, which should decrease the decoherence effect.

\section{Conclusions}
\label{sec:concl}

We have proposed two variations of the quantum-state transfer protocol in a spin chain that fulfill two basic requirements: simplicity and resistance to perturbation. They are immune to the inhomogeneity caused both by the local magnetic fields and disorder in exchange couplings and require only limited control of the external magnetic field at the terminal sites of the chain. Although the transfer time in the proposed protocol increases rapidly with the chain length, the fidelity of the transfer weakly depends on the chain length and it is possible to achieve state transfer on demand using a variable magnetic field (adiabatic protocol). This compares favorably with the earlier proposals. We have also shown that the amount of the information available in the central part of the chain is small during the whole transfer process and can be decreased even more by increasing the magnitude of the external magnetic field. This can be considered as an advantage, as it decreases the probability for the eavesdropper to distinguish between two orthogonal states sent and increases the time needed for the interception of the quantum state. However, it also affects the information transmission time, therefore a trade-off is expected between security and the transfer speed. Finally, let us note that the proposed adiabatic protocol allows a trivial extension to a dual-rail protocol \cite{burgarth05a}.
\\~~\\
\textbf{Acknowledgments:} This work was supported in part by the TEAM programme of the Foundation for Polish Science, cofinanced by the European Regional Development Fund (K.K. and P.M.). P.H. acknowledges partial support from the National Science Centre project 2011/01/B/ST2/05459.

\bibliographystyle{prsty}
\bibliography{abbr,quantum}

\end{document}